\documentclass[article]{IEEEtran}
\usepackage[intlimits]{amsmath}
\usepackage{amsfonts}
\usepackage{tikz}
\usepackage{cite}
 \usepackage[normalem]{ulem}
\usepackage{circuitikz}
 \usepackage{tkz-euclide}
 \usetikzlibrary{angles}
 \usetikzlibrary{positioning}
\usepackage{tikz-3dplot}
\usepackage[utf8]{inputenc}
\usepackage{color}
\usepackage{lipsum}
\usepackage{soul}
\usepackage{algorithm}
\usepackage{algpseudocode}
\usepackage{setspace}
\usepackage{float}
\usepackage{outlines}
\usepackage{physics}
\newfloat{algorithm}{t}{lop}
\usepackage{physics}
\usepackage{multicol}
\usepackage{tabularx}
\usepackage{booktabs}
\usepackage{url}

\newcommand{\M}[1]{\mathbf{#1}}
\newcommand{\T}[1]{\mathrm{#1}}

\newcommand{\eg}{\textit{e}.\textit{g}.{}}

\title{Signal Fidelity in Degenerate and Nondegenerate Mode Parametric Amplifier Receiving Antennas}
\author{Clayton Blosser, \IEEEmembership{Student Member, IEEE}, Adrian Bauer, \IEEEmembership{Student Member, IEEE}, Jessica E.~Ruyle, \IEEEmembership{Senior Member, IEEE}, K.C.~Kerby-Patel, \IEEEmembership{Senior Member, IEEE}, and Kurt~Schab, \IEEEmembership{Member, IEEE}

\thanks{Manuscript received \today; revised \today.}%
\thanks{C. Blosser, A. Bauer, and J. Ruyle are with the Advanced Radar Research Center at University of Oklahoma, Norman, USA, (e-mails: \{clayton.g.blosser, adrian.l.bauer-1,ruyle\}@ou.edu).}%
\thanks{K.C. Kerby-Patel is with University of Massachussetts Boston, Boston, USA (e-mail: kc.kerby-patel@umb.edu).}
\thanks{K. Schab is with Santa Clara University, Santa Clara, USA (e-mail: kschab@scu.edu).}%
}
\begin{document}
\let\Algorithm\algorithm
\renewcommand\algorithm[1][]{\Algorithm[#1]\setstretch{1.4}}

\pagestyle{empty}
\onecolumn

\setcounter{page}{1}

\newpage
\pagestyle{headings}
\twocolumn

\maketitle

\begin{abstract}
The gain, received power bandwidth, transient characteristics, and signal fidelity of two time-varying electrically small antennas based on parametric amplifier design are studied using practical QAM signals. Results show that interference from the difference harmonic present in the response of degenerate-mode parametric amplification decreases its signal throughput relative to a reference linear time-invariant (LTI) receiver, despite its apparent increased received power bandwidth in the frequency domain. The analysis also demonstrates that a non-degenerate parametric receiver, lacking this detrimental effect, exhibits increased signal throughput over the reference LTI receiver.
\end{abstract}


\section{Introduction}

\IEEEPARstart{P}{arametric} amplifiers (paramps) based on periodically time-varying reactive loads~\cite{manley1956some,blackwell1961} have a rich history in optics and high-frequency circuits, covering a range of applications including high-efficiency microwave amplifiers~\cite{gray2011broadband}, optical systems~\cite{liu2010mid,torounidis2006fiber}, up- and down-conversion for radar~\cite{li2009millimeter, hedayati2021parametric}, and amplification for quantum systems~\cite{yurke1989, abdo2011,roy2015broadband,white2015traveling}.  In many of these example areas, the low-noise characteristics of paramps are of prime importance.

Paramp-inspired time-varying antenna loading has previously been used to realize on-aperture low-noise amplification~\cite{frost1964Parametric, loghmannia2019active} and to circumvent the Bode-Fano bandwidth limit~\cite{loghmannia2021broadband, mekawy2021parametric} for high-$Q$, electrically small antenna (ESA) systems. Recent ESA-related paramp research has focused on phase-coherent degenerate-mode architectures because of their advantages over non-degnerate parametric amplifier architectures: they have larger bandwidth for a given variation in capacitance, are easier to match to standard line impedances, 
and require one fewer resonator~\cite{blackwell1961,loghmannia2019active,loghmannia2021broadband,mekawy2021parametric}. 

Despite these advantages, degenerate mode parametric amplification also has drawbacks compared to the non-degenerate case. Negative resistance parametric amplification, a feature common to both degenerate and non-degenerate systems studied here, produces a frequency-reversed copy of the input signal spectrum centered at an intermediate idler frequency~\cite[\S 3.3]{blackwell1961}. In degenerate paramp structures, this reversed spectrum appears at the output along with the desired signal spectrum. 
This effect can produce phase-dependent gain and interference between the signal and idler harmonics unless more complex paramp designs or detection techniques are employed~\cite[\S 6.3]{blackwell1961}. These effects have been observed in parametric amplifier circuits and can also be expected to occur in antennas co-designed with parametric amplification.

Recent antennas co-designed with degnerate mode parametric amplification have reported exclusively frequency-domain performance metrics, which cannot detect this signal-idler interference effect
\cite{loghmannia2019active, mekawy2021parametric, loghmannia2021broadband}.
This paper extends previous analyses by investigating the effect of degenerate-mode phase-dependent gain on the reception of modern communication schemes, such as quadrature amplitude modulation (QAM), which encode information in the signal phase.  
Results demonstrate that the presence of the difference harmonic in the degenerate-mode paramp output band corrupts the output symbol constellation and seriously degrades EVM performance relative to linear time-invariant (LTI) and non-degenerate paramp systems with ostensibly narrower bandwidths.

\section{Classes of Parametric Loading}\label{sec:designs}
Negative resistance paramps use a pumped nonlinear reactance to create gain at a signal frequency $f_\T{s}$~\cite{manley1956some, blackwell1961}. Amplification requires that the nonlinear device's reactance be canceled out at a passband resonance and at the idler frequency $f_\T{i}=\left|f_\T{s}-f_\T{p}\right|$ using external circuitry or resonating structures. Power is delivered to the nonlinear reactance at the pumping frequency $f_\T{p}$ by an external driving source. While parametric amplifiers can theoretically be implemented at any frequency using a wide variety of time-varying components or materials, 
for RF applications it is common to use a nonlinear capacitance realized through a varactor. The nonlinear capacitance can be modeled as a time-varying capacitance if the power delivered by the pumping source is much greater than the signal power, such that the varactor's response to the input signal is locally linear \cite[\S 3.4]{maas2003}. Under these assumptions, and using a sinusoidal pumping signal, the time-varying capacitance used throughout this letter takes the form
\begin{equation}
    C(t) = C_0(1+\gamma\sin(2\pi f_\T{p} t)),
\end{equation}
where $C_0$ is the static capacitance and $\gamma \in [0,1)$ is a modulation factor describing the depth of variation in capacitance.

There are two distinct classes of negative resistance paramps; non-degenerate paramps have a pumping frequency sufficiently high that the idler does not fall into the signal passband, while degenerate paramps select a pumping frequency such that the idler falls within the signal passband~\cite[\S 3]{blackwell1961}. Both types achieve gain at the signal frequency $f_\T{s}$ because power contributed at the signal frequency by the idler harmonic creates an apparent negative resistance. If the pump signal of a degenerate paramp is twice the passband's resonant frequency and its idler and signal are phase-synchronous, it is phase-coherent and results in a higher output power over a wider bandwidth relative to non-degenerate paramp. 
Additionally, the degenerate paramp has a larger negative resistance for a given modulation factor, allowing a resonator with a small resistance to be matched to a much higher line impedance. Time-varying antenna designs utilizing non-degenerate and phase-coherent degenerate parametric amplification are studied in detail in Section~\ref{sec:arch}.

\section{Antenna Architecture}\label{sec:arch}
\begin{figure}
    \centering
    \input{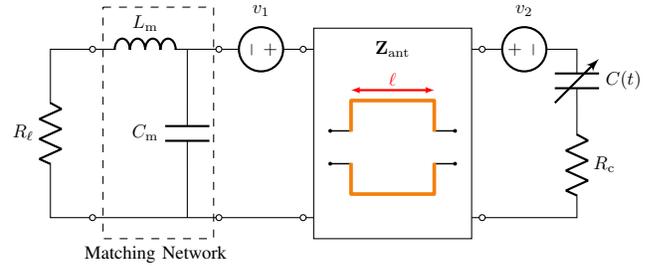}
    \caption{Loop antenna and incident plane wave modeled as a two-port network $\M{Z}_\T{ant}$ and voltage sources $v_1$ and $v_2$ with L matching network.  For the LTI design, $C(t)$ is static and $R_\T{c}=0$.}
    \label{fig:diag}
\end{figure}

Throughout the remainder of this letter, we study three loading configurations applied to a two-port electrically small square planar loop antenna similiar to the transmitting antenna design reported in \cite{mekawy2021parametric}. Here we consider its receiving properties to study signal fidelity effects. It has the approximately uniform current distribution typical of this antenna type~\cite[\S 5.2]{Balanis1989} with a 3~mm trace width and 150~mm outer side length, see Fig.~\ref{fig:diag}. Ports are located at the center of two opposite sides of the loop. The antenna is tuned to a first resonance at $100$~MHz, which is used as the operating frequency. The loop's electrical size $ka$ is $0.22$. The trace conductivity is that of copper ($\sigma = 5.7\cdot10^7$~S/m).

To efficiently extract the receiving properties of different loading configurations, the loop structure was modeled with method of moments (MoM)~\cite{atom} and the resulting impedance matrix was compressed to a two-port network using a Schur complement method. A similar compression was applied to the voltage excitation vector associated with a co-polarized plane wave of unit magnitude incident on the loop's broadside direction. Using MoM data collected across broad bandwidths, we model each candidate loading configuration using the equivalent circuit shown in Fig.~\ref{fig:diag}. There, $\M{Z}_\T{ant}$ is the two-port broadband compressed impedance matrix of the loop structure, $\M{V} = [v_1, v_2]$ is the compressed excitation vector, $L_\T{m}$ and $C_\T{m}$ constitute a matching network (MN), $R_\ell$ is a load impedance, and the combination of $C(t)$ and $R_\T{c}$ model a lossy, time-varying capacitance~\cite{varactor2020}.  The frequency-domain and transient response of this circuit under varying loading conditions is then simulated using three independent methods: in-house conversion matrix method of moments (CMMoM)~\cite{bass2022conversion}, commercial harmonic balance (HB), and commercial transient circuit co-simulation \cite{ADS}. Frequency bounds on initial MoM calculations are selected to obtain stable responses in both transient and frequency domain calculations involving high-order harmonics of the pumping and driving signal.

Three loading configurations are studied, each designed to balance received power gain and bandwidth while matched to a load impedance of $R_\ell = 50~\Omega$. Parameter values for the three loading configurations are shown in Table \ref{tab:my_label}. Capacitance, modulation factor, and equivalent series loss resistance for the time-varying capacitor in the degenerate time-varying (DTV) and non-degenerate time-varying (NDTV) cases were chosen based on a Skyworks SMV1234 varactor~\cite{varactor2020} and modeled using the SPICE model in \cite{Fisher1986Modeling}. The static capacitance in the LTI design was assumed to be lossless. Modulation factors for both non-LTI designs are equal to provide a fair comparison. The DTV design was tuned to the load impedance via the paramp, omitting the circuit in the dashed box in Figure~\ref{fig:diag}. The LTI and NDTV designs require an external L-MN~\cite[\S 5.1]{pozarMicrowaveEngineering2011}. All MN inductors are lossy with a $Q=100$.

\def\arraystretch{1.1}
\begin{table}[]
\setlength{\tabcolsep}{4pt} 
\renewcommand{\arraystretch}{1} 
    \centering
    \begin{tabular}{c|c|c|c|c}
        Config. & $L_\T{m}$/$C_\T{m}$  (nH/pF)&
        $C_0$ (pF) & $\gamma$ & $f_\T{p}$ (MHz)\\\hline
        LTI & 9 / 277 & 4.0  & 0 & --\\
        DTV  & -- & 4.1  & 0.332 & 200\\
        NDTV & 27 / 95 & 4.2  & 0.332 & 669\\
    \end{tabular}
    \vspace{0.1in}
    \caption{Specification of three loading configurations.  
    }
    \label{tab:my_label}
\end{table}

\begin{figure}[tbp]
    \centering
    \includegraphics[width=0.4\textwidth]{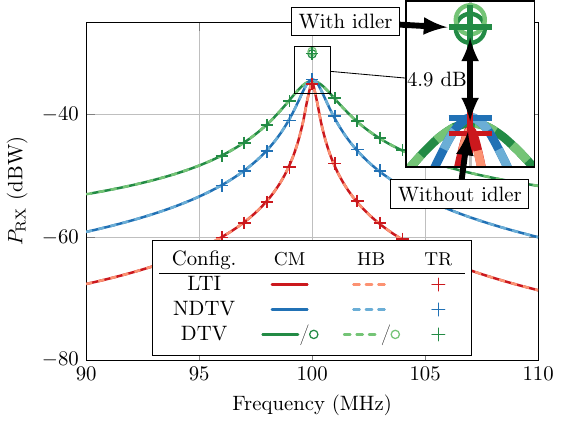} 
    \caption{Received power as a function of frequency, computed using CMMoM, HB, and transient methods. Markers for CMMoM and HB simulations indicate power at 100~MHz for the DTV design with power from the idler harmonic included.}
    \label{fig:p-vs-f}
\end{figure}

Figure~\ref{fig:p-vs-f} shows the power received by each design as a function of incident frequency when excited by a unit-amplitude ($1~\T{V}/\T{m}$), copolarized plane wave with time dependence $\cos(\omega t + \phi)$ for a single phase $\phi$ relative to the pumping signal at frequency $\omega_\T{p}$. The difference in the three simulation methods, CMMoM, HB, and transient, is under 0.5 dB at 100 MHz, and under 0.1 dB elsewhere.  The peak LTI received power is $-34.5$~dBW at the center frequency, which agrees with the expected realized effective aperture of a perfectly matched small loop with peak directivity $D\approx1.5$ and total efficiency $26.8\%$\footnote{The relatively low net efficiency of 26.8\% is primarily due to the  29.1\% radiation efficiency of the antenna since the MN's efficiency is 92.0\%} (including antenna and MN losses) obtained from MoM calculations \cite[\S 5.2]{Balanis1989}, \cite{best2016optimizing}. In the DTV case, there is an isolated spike of 4.9 dB in received power at 100 MHz due to constructive interference between the signal and idler harmonics. The frequency-domain 
methods do not immediately capture this effect because they compute the power received at each harmonic separately. Adding the idler and signal frequency contributions eliminates this discrepancy between the transient and frequency-domain methods.

The $-3$~dB fractional bandwidths of the  LTI, NDTV, and DTV receivers are $0.4\%$, $1.1\%$, and $2.0\%$, respectively. The fractional bandwidth of the DTV design was calculated neglecting the 4.9 dB spike that occurs when the signal and idler constructively interfere. The NDTV and DTV designs' bandwidths were greater than that of the LTI design by factors of 2.5 and 4.7, respectively. Using the same modulation factor, the degenerate receiver was capable of almost twice the bandwidth as the non-degenerate and also did not require a MN.  Based on the data in Fig.~\ref{fig:p-vs-f} alone, one would expect the DTV to have the highest information throughput of the three designs. Next, Section \ref{sec:si-fi} examines the impact of signal-idler interaction in the degenerate mode design on signal fidelity.

\section{Signal fidelity analysis}
\label{sec:si-fi}
We now consider the ability of each system to receive QAM-like input signals.  To this end, we characterize each design by its response to a series of unit-amplitude, co-polarized plane wave excitations with time dependence $U(t)\cos(\omega t + \phi)$, where $U(t)$ is the Heaviside step function.  These responses are computed using the same transient co-simulation method used to generate data in Fig.~\ref{fig:p-vs-f}.  The resulting RF waveforms are then downconverted to their complex baseband representations $y(t,\phi)$ for various phases $\phi$.

\begin{figure}
    \centering
    \includegraphics[width=3.25in]{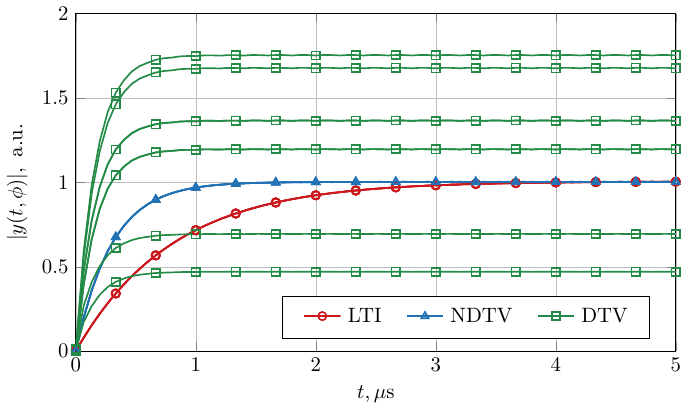}
    \caption{Magnitudes of baseband step responses $y(t,\phi)$ from each design under varying input signal phases, relative to the pump signal.  All incident phases lead to identical LTI and NDTV responses, while the DTV system has a phase-dependent characteristic, leading to multiple unique traces.}
    \label{fig:steps}
\end{figure}

The absolute values of these responses are shown in Fig.~\ref{fig:steps} for several incident phases.  As expected, the LTI and NDTV designs show no phase dependence, with all incident phases producing identical responses.  In contrast, the DTV loading scheme produces responses that vary significantly in amplitude as the incident phase is changed relative to the pump signal due to interaction with the idler harmonic, as evidenced by multiple DTV traces. This phase dependence can degrade information encoded via a received signal's amplitude and phase (\eg, QAM-like modulation). The rise time of each system's response is roughly inversely proportional to the received power bandwidth observed in Fig.~\ref{fig:p-vs-f}.  

\begin{figure}
    \centering
    \includegraphics[height=1.2in]{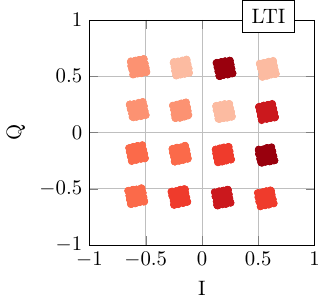}
    \includegraphics[height=1.2in]{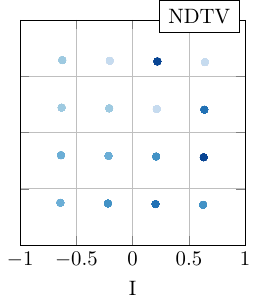}
    \includegraphics[height=1.2in]{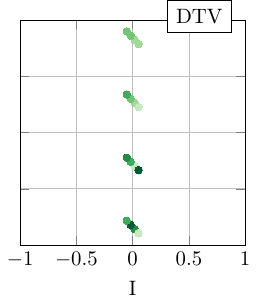}
    \caption{Noise-free, unequalized constellations from each system using a 2048-symbol 16QAM PRBS at 0.5~Msym/s.}
    \label{fig:const}
\end{figure}

To demonstrate the detrimental effect of the DTV system's phase-dependency, we compute the baseband response of each design when driven with a pseudorandom bit sequence (PRBS) of symbols using a 16QAM constellation.  To avoid lengthy transient simulations of this entire PRBS at RF-scale time resolution, we directly compute the entire complex baseband response of each system via convolution of the complex symbol sequence with individual step responses shown in Fig.~\ref{fig:steps}.  Specifically, we represent the incident RF signal $s(t)$ as a series of weighted, time-delayed, and phase-shifted modulated step functions corresponding to the $N$ symbols of the PRBS,
\begin{multline}
 s(t) = \sum_{n=1}^{N} a_nU(t-\tau_n) \cos(\omega (t-\tau_n) + \phi_n)\\ - a_{n-1}U(t-\tau_{n}) \cos(\omega (t-\tau_{n}) + \phi_{n-1})
\end{multline}
where $a_n$, $\phi_n$, and $\tau_n$ are the amplitude, phase, and start time of the $n^\T{th}$ symbol, and $a_0=0$. 
The system's complex baseband response $y_s(t)$ to the excitation $s(t)$ is constructed by a weighted superposition of its baseband step responses,
\begin{equation}
    y_s(t) = \sum_{n=1}^{N} a_n y(t-\tau_n,\phi_n) - a_{n-1}y(t-\tau_{n},\phi_{n-1})
\end{equation}
where each input term $U(t-\tau_n) \cos(\omega (t-\tau_n) + \phi_n)$ is mapped to the output response $y(t-\tau_n,\phi_n)$.
Note that this formulation implies sampling (or interpolation) of each system's step responses for all possible symbol phases $\{\phi_n\}$ included in the selected constellation.

Output constellations from each system for a 2048-symbol PRBS at 0.5~Msym/s are shown in Fig.~\ref{fig:const}.  The baseband waveform is noise-free and all spread observed in constellation points is due solely to intersymbol interference (ISI) effects.  In agreement with expectations from Figs.~\ref{fig:p-vs-f} and \ref{fig:steps}, the LTI and NDTV systems reproduce the incident 16QAM constellation with varying degrees of ISI, whereas the DTV system's constellation is highly compressed along one dimension due to its phase-dependent response.  

\begin{figure}
    \centering
\includegraphics[width=3.25in]{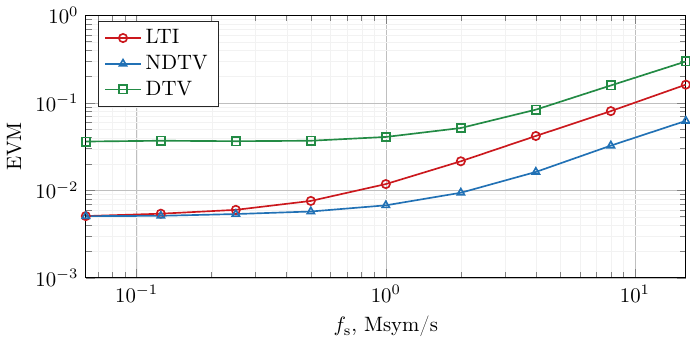}
    \caption{Post-equalization EVM for each system at varying data rates in the presence of AWGN.  A constant internal noise temperature was used across each data rate and system, with the LTI system showing SNR~$\approx$~30~dB at 0.25~Msym/s.}
    \label{fig:evm}
\end{figure}

Though each constellation can be equalized prior to symbol estimation, the efficacy of such equalization in the presence of noise is greatly reduced by ISI and other forms of distortion.  This effect is demonstrated in Fig.~\ref{fig:evm}, where the error vector magnitude (EVM) of equalized constellations from each system is shown as a function of incident data rate.  No external noise was included, but a consistent internal noise temperature was used for each system, resulting in a signal to noise ratio (SNR) of approximately 30~dB at 0.25~Msym/s from the LTI system.  In all cases, the signal power scales linearly with data rate to maintain a constant energy per symbol.  Identical 128-tap, 16-to-1 downsampling, linear equalizers were trained until convergence on noisy instances of a 2048-symbol PRBS for each system and data rate prior to recording EVM.

Each system exhibits similar trends across data rates, with converging EVM at low data rates (noise-limited regime) and an upward trend at high data rates (bandwidth-limited regime).  The LTI and NDTV systems show similar noise-limited characteristics due to comparable gains and relatively predictable ISI, though, consistent with results in Figs.~\ref{fig:p-vs-f}-\ref{fig:const}, the NDTV enters the bandwidth-limited regime at approximately 3$\times$ higher data rates than the LTI system owing to its broader received power bandwidth.  The DTV system, however, shows significantly higher EVM in the noise-limited regime due to its compressed pre-equalized constellation, which greatly increases the effect of added noise. Though the gap between the DTV and the other two systems narrows in the bandwidth-limited regime, the DTV receiver consistently produces higher EVM across all data rates.

\section{Discussion}\label{sec:discussion}
Results presented in Section~\ref{sec:si-fi} showed that the idler harmonic in the output band of the degenerate mode parametric receiver degrades the EVM of phase-dependent communication schemes. Due to the frequency scalability of paramps, this assessment holds at varying frequencies of operation. This result was demonstrated using QAM signals and additionally applies to any system that utilizes QAM for spreading (DSSS) or on subcarriers (OFDM). As a result, non-degenerate mode parametric amplification is preferable over degenerate mode parametric amplification in receiving systems unless a degenerate-compatible modulation scheme is used or alterations to the paramp architecture are made. One alternative is a phase-independent modulation scheme such as on-off keying or amplitude shift keying, 
but this requires precise phase-locking 
to prevent phase-dependent attenuation. Single-sideband modulation could also preserve signal fidelity
, but at the cost of half the apparent bandwidth of the DTV system. 

\begin{figure}
    \centering
    \includegraphics[width=3.25in]{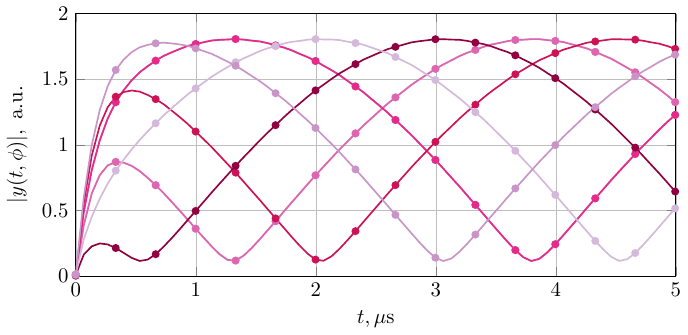}
    \caption{Magnitudes of baseband step responses $y(t,\phi)$ for a 100~MHz DTV system with incident frequency and downconversion process detuned to 100.1~MHz.  The exact time-variation is phase-dependent, as illustrated by multiple traces, but the time-average received power is phase-independent.}
    \label{fig:beats}
\end{figure}

The degenerate paramp architecture may also be modified to mitigate the observed signal fidelity issues. A phase-incoherent degenerate mode parametric receiver could be employed by operating the pump frequency at a slightly higher or lower frequency than the incident carrier~\cite[\S 3.3]{blackwell1961}. Relocating the pumping frequency does not unconditionally prevent self-interference, because the frequency-reversed copy of the signal still appears in the passband. Even if interference is avoided with sufficient spacing between the signal and idler spectra, a beat frequency is produced in the baseband response, as shown in Fig.~\ref{fig:beats}, which requires sharp filtering to remove and otherwise significantly degrades signal fidelity. Alternatively, one harmonic can be canceled from the output using a balanced pair of degenerate mode parametric amplifiers with pump signals exactly $180^\circ$ out of phase~\cite[\S 6.3]{blackwell1961}. This and other balanced architectures permit use of the entire bandwidth.

Future work on degenerate mode parametric amplifying antennas, particularly in receive mode, should employ the aforementioned techniques to avoid the phase-dependent signal degradation illustrated in this paper.  At a minimum, the signal fidelity of proposed degenerate mode receiving systems should be studied carefully, as the results in this work demonstrate that classical frequency domain quantities (bandwidth, received power) may not fully represent the efficacy of such receivers within practical communication systems.

\section*{Acknowledgment}
This research is based upon work supported in part by the Office of the Director of National Intelligence (ODNI), Intelligence Advanced Research Projects Activity (IARPA), via [2021-2106240007]. The views and conclusions contained herein are those of the authors and should not be interpreted as necessarily representing the official policies, either expressed or implied, of ODNI, IARPA, or the U.S. Government. The U.S. Government is authorized to reproduce and distribute reprints for governmental purposes notwithstanding any copyright annotation therein.

\bibliographystyle{ieeetr}
\bibliography{references,nLTIrefs}

\begin{thebibliography}{10}

\bibitem{manley1956some}
J.~Manley and H.~Rowe, ``Some general properties of nonlinear elements-{Part} {I}. {G}eneral energy relations,'' {\em Proceedings of the IRE}, vol.~44, no.~7, pp.~904--913, 1956.

\bibitem{blackwell1961}
L.~A. Blackwell and K.~L. Kotzebue, {\em Semiconductor-Diode Parametric Amplifiers}.
\newblock Englewood Cliffs, N.J.: Prentice-Hall, 1961.

\bibitem{gray2011broadband}
B.~Gray, F.~Ramirez, B.~Melville, A.~Suarez, and J.~S. Kenney, ``A broadband double-balanced phase-coherent degenerate parametric amplifier,'' {\em IEEE Microwave and Wireless Components Letters}, vol.~21, no.~11, pp.~607--609, 2011.

\bibitem{liu2010mid}
X.~Liu, R.~M. Osgood, Y.~A. Vlasov, and W.~M. Green, ``Mid-infrared optical parametric amplifier using silicon nanophotonic waveguides,'' {\em Nature Photonics}, vol.~4, no.~8, pp.~557--560, 2010.

\bibitem{torounidis2006fiber}
T.~Torounidis, P.~A. Andrekson, and B.-E. Olsson, ``Fiber-optical parametric amplifier with 70-{dB} gain,'' {\em IEEE Photonics Technology Letters}, vol.~18, no.~10, pp.~1194--1196, 2006.

\bibitem{li2009millimeter}
J.~Li, Y.~Liang, and K.~K.-Y. Wong, ``Millimeter-wave {UWB} signal generation via frequency up-conversion using fiber optical parametric amplifier,'' {\em IEEE Photonics Technology Letters}, vol.~21, no.~17, pp.~1172--1174, 2009.

\bibitem{hedayati2021parametric}
M.~Hedayati, L.~K. Yeung, M.~Panahi, X.~Zou, and Y.~E. Wang, ``Parametric downconverter for mixer-first receiver front ends,'' {\em IEEE Transactions on Microwave Theory and Techniques}, vol.~69, no.~5, pp.~2712--2721, 2021.

\bibitem{yurke1989}
B.~Yurke, L.~R. Corruccini, P.~G. Kaminsky, L.~W. Rupp, A.~D. Smith, A.~H. Silver, R.~W. Simon, and E.~A. Whittaker, ``Observation of parametric amplification and deamplification in a {Josephson} parametric amplifier,'' {\em Phys. Rev. A}, vol.~39, pp.~2519--2533, Mar 1989.

\bibitem{abdo2011}
B.~Abdo, F.~Schackert, M.~Hatridge, C.~Rigetti, and M.~Devoret, ``{Josephson amplifier for qubit readout},'' {\em Applied Physics Letters}, vol.~99, p.~162506, 10 2011.

\bibitem{roy2015broadband}
T.~Roy, S.~Kundu, M.~Chand, A.~Vadiraj, A.~Ranadive, N.~Nehra, M.~P. Patankar, J.~Aumentado, A.~Clerk, and R.~Vijay, ``Broadband parametric amplification with impedance engineering: Beyond the gain-bandwidth product,'' {\em Applied Physics Letters}, vol.~107, no.~26, 2015.

\bibitem{white2015traveling}
T.~White, J.~Mutus, I.-C. Hoi, R.~Barends, B.~Campbell, Y.~Chen, Z.~Chen, B.~Chiaro, A.~Dunsworth, E.~Jeffrey, {\em et~al.}, ``Traveling wave parametric amplifier with {Josephson} junctions using minimal resonator phase matching,'' {\em Applied Physics Letters}, vol.~106, no.~24, 2015.

\bibitem{frost1964Parametric}
A.~Frost, ``Parametric amplifier antenna,'' {\em IEEE Transactions on Antennas and Propagation}, vol.~12, no.~2, pp.~234--235.

\bibitem{loghmannia2019active}
P.~Loghmannia and M.~Manteghi, ``An active cavity-backed slot antenna based on a parametric amplifier,'' {\em IEEE Transactions on Antennas and Propagation}, vol.~67, no.~10, pp.~6325--6333, 2019.

\bibitem{loghmannia2021broadband}
P.~Loghmannia and M.~Manteghi, ``Broadband parametric impedance matching for small antennas using the {B}ode-{F}ano limit: Improving on {C}hu’s limit for loaded small antennas,'' {\em IEEE Antennas and Propagation Magazine}, vol.~64, no.~5, pp.~55--68, 2021.

\bibitem{mekawy2021parametric}
A.~Mekawy, H.~Li, Y.~Radi, and A.~Al{\`u}, ``Parametric enhancement of radiation from electrically small antennas,'' {\em Physical Review Applied}, vol.~15, no.~5, p.~054063, 2021.

\bibitem{maas2003}
S.~A. Maas, {\em Nonlinear Microwave and RF Circuits}.
\newblock Norwood, MA: Artech House, 2nd~ed., 2003.

\bibitem{Balanis1989}
C.~A. Balanis, {\em Advanced Engineering Electromagnetics}.
\newblock Hoboken, NJ: Wiley, 1989.

\bibitem{atom}
{Czech Technical University in Prague}, ``Antenna toolbox for {MATLAB} {(AToM)}.'' \url{http://www.antennatoolbox.com/atom}.

\bibitem{varactor2020}
Skyworks, ``{Skyworks'} {SMV123XXX} series varactors.'' \url{https://www.mouser.com/datasheet/2/472/SMV123x\_\linebreak Series\_200058AA-3364699.pdf}, 2020.
\newblock Accessed: (01/10/2024).

\bibitem{bass2022conversion}
S.~F. Bass, A.~M. Palmer, K.~R. Schab, K.~C. Kerby-Patel, and J.~E. Ruyle, ``Conversion matrix method of moments for time-varying electromagnetic analysis,'' {\em IEEE Transactions on Antennas and Propagation}, vol.~70, no.~8, pp.~6763--6774, 2022.

\bibitem{ADS}
{Keysight}, ``Pathwave advanced design system.'' \url{https://www.keysight.com/us/en/products/software/pathwave-design-software/pathwave-advanced-design-system.html}.

\bibitem{Fisher1986Modeling}
G.~Fisher and J.~Connelly, ``Modeling time-dependent elements for {SPICE} transient analyses,'' {\em IEEE Transactions on Computer-Aided Design of Integrated Circuits and Systems}, vol.~5, no.~3, pp.~429--432, 1986.

\bibitem{pozarMicrowaveEngineering2011}
D.~Pozar, {\em Microwave {{Engineering}}}.
\newblock John Wiley \& Sons, Inc.

\bibitem{best2016optimizing}
S.~R. Best, ``Optimizing the receiving properties of electrically small {HF} antennas,'' {\em URSI Radio Science Bulletin}, vol.~2016, no.~359, pp.~13--29, 2016.

\end{thebibliography}

\end{document}